\documentclass[conference]{IEEEtran}
\IEEEoverridecommandlockouts
\usepackage{cite}
\usepackage{amsmath,amssymb,amsfonts}
\usepackage{algorithmic}
\usepackage{graphicx}
\usepackage{textcomp}
\usepackage{xcolor}
\usepackage{subcaption}
\usepackage{tikz}
\usetikzlibrary{quantikz}

\newtheorem{approximation}{Approximation}

\def\BibTeX{{\rm B\kern-.05em{\sc i\kern-.025em b}\kern-.08em
    T\kern-.1667em\lower.7ex\hbox{E}\kern-.125emX}}
\begin{document}

\title{Compiler for Distributed Quantum Computing: \\ a Reinforcement Learning Approach
}

 \author{
      \IEEEauthorblockN{Panagiotis Promponas\IEEEauthorrefmark{1}\IEEEauthorrefmark{2}\IEEEauthorrefmark{3}, Akrit Mudvari\IEEEauthorrefmark{1}\IEEEauthorrefmark{3}\thanks{\IEEEauthorrefmark{3} These authors contributed equally to this work.}, Luca Della Chiesa\IEEEauthorrefmark{2}, Paul Polakos\IEEEauthorrefmark{2},
      Louis Samuel\IEEEauthorrefmark{2},
      Leandros Tassiulas\IEEEauthorrefmark{1}\thanks{Correspondence to \texttt{panagiotis.promponas@yale.edu}} }
      \IEEEauthorblockA{\IEEEauthorrefmark{1}Department of Electrical Engineering, Yale University
     }
      \IEEEauthorblockA{\IEEEauthorrefmark{2}Cisco Systems
      }
  }

\maketitle

\begin{abstract}
The practical realization of quantum programs that require large-scale qubit systems is hindered by current technological limitations. Distributed Quantum Computing (DQC) presents a viable path to scalability by interconnecting multiple Quantum Processing Units (QPUs) through quantum links, facilitating the distributed execution of quantum circuits. In DQC, EPR pairs are generated and shared between distant QPUs, which enables quantum teleportation and facilitates the seamless execution of circuits. A primary obstacle in DQC is the efficient mapping and routing of logical qubits to physical qubits across different QPUs, necessitating sophisticated strategies to overcome hardware constraints and optimize communication. We introduce a novel compiler that, unlike existing approaches,  prioritizes reducing the expected execution time by jointly managing the generation and routing of EPR pairs, scheduling remote operations, and injecting SWAP gates to facilitate the execution of local gates. We present a real-time, adaptive approach to compiler design, accounting for the stochastic nature of entanglement generation and the operational demands of quantum circuits. Our contributions are twofold: (i) we model the optimal compiler for DQC using a Markov Decision Process (MDP) formulation, establishing the existence of an optimal algorithm, and (ii) we introduce a constrained Reinforcement Learning (RL) method to approximate this optimal compiler, tailored to the complexities of DQC environments. Our simulations demonstrate that Double Deep Q-Networks (DDQNs) are effective in learning policies that minimize the depth of the compiled circuit, leading to a lower expected execution time and likelihood of successful operation before qubits decohere. 
 
\end{abstract}


 



\section{Introduction}

Quantum computing is set to revolutionize problem-solving capabilities beyond classical computers' limits, utilizing algorithms like Shor's \cite{Sho99}. However, substantial quantum computers require thousands of qubits, a goal yet unmet by current models \cite{ibmroadmap, ibm127, arute2019quantum}. Distributed Quantum Computing (DQC) offers a solution by networking smaller, manageable \textit{Quantum Processing Units (QPUs)} to operate as a cohesive unit \cite{saleem2021quantum, cirac1999distributed, gyongyosi2021scalable, MD16, guha2022cluster, qiao2022quantum, cacciapuoti2019quantum}. This approach, conceptualized by Grover \cite{grover1997quantum} and Cleve and Buhrman \cite{cleve1997substituting}, facilitates the distributed execution of quantum circuits across these processors, despite their physical separation. Critical to DQC is quantum teleportation, which allows for the transfer of qubits and gates between processors, overcoming physical qubit interaction limits \cite{barral2024review}. The challenge lies in quantum compilation, adapting theoretical circuits to the constraints of actual quantum hardware, especially in DQC where entanglement control and operational scheduling are pivotal. See \cite{barral2024review} for a detailed review of DQC.

Given a quantum circuit, there are two necessary procedures to effectively execute it in a DQC environment; initial qubit mapping, and qubit routing.\footnote{Both of these procedures have a (simplified) counterpart in quantum compilation of a single QPU (e.g., \cite{botea2018complexity,moro2021quantum,zhu2020exact, ash2019qure, finigan2018qubit,pozzi2022using}).} During the initial qubit mapping phase, the goal is to map the logical qubits of the circuit to the physical qubit memories within the DQC architecture. Given multiple interconnected QPUs, the decision involves determining which qubits will be allocated to QPUs that may be apart. After receiving the initial qubit mapping, the DQC compiler should implement the qubit routing, which (i) injects SWAP operations to enable gate execution on the actual hardware, (ii) generates EPR pairs in an optimized manner to cascade into the QPUs, and, (iii) inserts modules (e.g., gate and qubit teleportations) that facilitate interactions involving qubits separated across different QPUs.

Assuming only gate teleportations as a means towards DQC, previous works have used various heuristic methods reducing the problem of initial qubit mapping to \textit{graph partitioning} problems. The end goal is to minimize the number of non-local operations within a circuit assuming the bottleneck is the network communications \cite{davarzani2020dynamic,daei2020optimized, mao2023qubit}. These qubit partitioning approaches transform a circuit into a static qubit interaction graph for partitioning. Although these methods aim to group more frequently interacting qubits in the same partition, their efficiency is compromised by not considering the benefits of teleporting qubits to different QPUs when this could increase the number of gates that can be implemented locally.

Numerous papers also consider qubit teleportation to minimize the  communication cost (i.e., the number of EPR pairs needed for the execution of the quantum circuit) \cite{nikahd2021automated, davis2023towards, bandic2023mapping, ovide2023mapping, pastor2024circuit, chen2023routing, escofet2024revisiting, escofet2023hungarian}. These approaches strategize the mapping of qubits across various QPUs and consider the potential benefits of teleporting qubits to different QPUs to execute some of the remaining gates more efficiently, reducing communication costs. In \cite{andres2019automated, g2021efficient, sundaram2022distribution}, the authors consider solely cat-entanglement operations \cite{eisert2000optimal} as the means towards DQC to minimize the number of EPR pairs needed for the execution of a quantum circuit.

All of the aforementioned papers do not take into consideration the \textit{qubit routing} problem. Specifically, they do not optimize jointly the (i) generation and routing of the EPR pairs, (ii) scheduling of the remote operations (e.g., qubit or gate teleportations), and, (iii) compilation of the QPUs' ''local circuits". Most of the papers in the literature assume that the EPR pair generation is the only costly operation in the DQC environment, neglecting the \emph{state} of the DQC environment (e.g., the instantaneous position of the qubits and the congestion of the network that generates the EPR pairs). However, we propose that depending on the state of the QPUs and the remaining tasks of the circuit, implementing sparse (but a large number of) remote gates\footnote{Gates involving qubits across different QPUs.} could be more feasible/efficient than responding to frequent, small-scale demands for EPR pair generation using the quantum links (or quantum network).

In \cite{ferrari2023modular} the authors incorporate initial qubit mapping, remote gate scheduling, and qubit routing\footnote{This paper separates the remote gate scheduling from the qubit routing procedure, resulting in differing terminology than ours.} into the compilation process.
This work employs a k-partitioning heuristic for initial qubit mapping and uses a heuristic cost approach for scheduling remote gates, assessing the effectiveness of teleporting qubits versus gates. However, this cost heuristic is based solely on the count of future gates involving the same qubits, overlooking the potential need to move these qubits to different QPUs in the interim. Additionally, by distinguishing remote gate scheduling from qubit routing—a distinction our approach does not make since we jointly optimize them—it neglects to manage/optimize EPR pair generation, or to dynamically adjust to the availability of entanglements. This separation underlines a broader issue: decoupling qubit routing from the compilation process can limit the ability to respond adaptively to changes in the DQC environment's state. Finally, \cite{ferrari2021compiler} derives upper bounds of the overhead induced by quantum compilation for DQC.

\subsection{Contributions}

In this paper, we introduce a compiler model for gate-based DQC environments designed to minimize the execution time of quantum circuits by jointly optimizing the (i) generation and routing of EPR pairs, (ii) scheduling of the remote operations, and, (iii) injection of SWAP gates to facilitate the execution of local gates.
Moreover, using a similar technique with \cite{li2019tackling}, we can use this compiler model to optimize the initial qubit mapping (see Section~\ref{sec:discussion_and_model_extension}). Therefore, our proposed model accepts a quantum circuit as input and manages all necessary adaptations for seamless execution within a DQC environment. We argue that the optimal compiler should have the following characteristics (extending the list in \cite{ferrari2021compiler}):

\textit{General purpose:} The compiler should work for any given input circuit within a specified DQC environment.

\textit{Online:} The compiler should adapt to the stochastic nature of the system, which stems from the probabilistic nature of entanglement generation.

\textit{Efficient:} The compiler should operate efficiently due to the fragility of the qubits. As it also needs to function online, we propose that a trained model is suitable, allowing for extensive training time while ensuring that the actual compilation process is brief.

\textit{Effective:} The compiler should maximize the probability of successfully executing the circuit. For simplicity, we assume that this translates to minimizing the expected time required for the execution of the circuit. In Section~\ref{sec:discussion_and_model_extension} we discuss how we can incorporate heterogeneous gate errors into our model.


\textit{State-dependent:} The compiler should consider the whole state of the DQC environment to optimize its decisions. Consequently, two benefits of that would be that the compiler could (i) prepare/generate EPR pairs in advance and store them in the quantum memories within the QPUs to reduce the time required for executing future remote gates, and, (ii) accurately determine when to teleport a qubit or a gate by considering anticipated interactions across all qubits and their locations.

This paper makes the following contributions:

\begin{itemize}
    \item We model the optimal compiler for a DQC environment using an MDP formulation, guided by the five aforementioned characteristics. This model provides insight into its functionality and confirms that algorithms such as value iteration or policy iteration could potentially solve the MDP, ensuring the existence of an optimal solution for constructing a compiler in a DQC environment.

    \item  We propose a constrained RL model designed to effectively approximate the optimal policy for the compiler, efficiently handling the extensive state and action spaces. This method focuses on only the most essential environment information and uses heuristic reward-shaping to efficiently guide the RL agent towards optimal actions.

    
    \item We present simulation results showcasing the effectiveness of our RL approach, selected through a thorough investigation of various on and off-policy RL methods, in developing policies that reduce execution time and enhance the success rate of random quantum circuits.
\end{itemize}


To the best of our knowledge, this is the first compiler for a DQC environment designed to minimize \emph{expected} execution time. It integrates enhancements to the entanglement distribution network, improving EPR pair routing, remote operation scheduling, and strategic SWAP gate injection to support local gate execution. Unlike existing approaches in DQC where the optimization objective does not explicitly consider the \emph{actual} elapsed time (measured in terms of CNOT gate operations or concurrent gate executions), our compiler specifically aims to minimize the real-time duration required for circuit execution. Finally, unlike the aforementioned works, our trained model is capable of compiling circuits on-the-fly, eliminating the need to run an algorithm for each new circuit.

\section{Preliminaries}
\label{sec:preliminaries}
This section introduces the preliminaries and the notation used by summarizing the notions of quantum circuits and quantum teleportation (Section~\ref{ssec:quantum_circuits_and_entanglement}), and  describing QPU and DQC architectures (Sections~\ref{ssec:QPU_architecture}~and~\ref{ssec:DQC_architecture} respectively).

\subsection{Quantum Gates \& Quantum Teleportation}
\label{ssec:quantum_circuits_and_entanglement} 

\begin{figure}[ht]
    \centering
    \begin{quantikz}[row sep=0.04cm]
        \lstick{$q_0$} & \ctrl{1} & \ctrl{2} & \qw      & \ctrl{4} & \qw      & \qw      & \ctrl{2} & \ctrl{1} & \qw \\
        \lstick{$q_1$} & \targ{}  & \qw      & \ctrl{4} & \qw      & \ctrl{5} & \qw      & \qw      & \targ{}  & \qw \\
        \lstick{$q_2$} & \qw      & \targ{}  & \qw      & \qw      & \qw      & \ctrl{1} & \targ{}  & \qw      & \qw \\
        \lstick{$q_3$} & \qw      & \qw      & \qw      & \qw      & \qw      & \targ{}  & \qw      & \qw      & \qw \\
        \lstick{$q_4$} & \qw      & \qw      & \qw      & \targ{}  & \qw      & \qw      & \qw      & \qw      & \qw \\
        \lstick{$q_5$} & \qw      & \qw      & \targ{}  & \qw      & \qw      & \qw      & \qw      & \qw      & \qw \\
        \lstick{$q_6$} & \qw      & \qw      & \qw      & \qw      & \targ{}  & \qw      & \qw      & \qw      & \qw \\
    \end{quantikz}
    \vspace{-0.3cm}
    \caption{A circuit comprising 7 qubits and exclusively CNOT operations as specified.}
    \label{fig:quantum_circuit}
\end{figure}
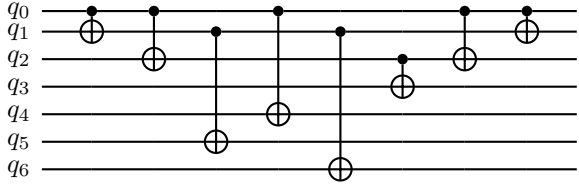

Gate-based quantum computation relies on a sequence of controlled operations, such as the Controlled-NOT (CNOT) gates. Using 3 CNOT gates, a SWAP gate can be implemented, which exchanges the states of two qubits without altering their individual states. SWAP gates will be important for qubit routing, which is one of the main tasks of this paper. A CNOT gate along with any set of single-qubit gates constitutes a universal set for quantum computation, allowing any unitary operation to be approximated with arbitrary accuracy \cite{nielsen2010quantum, barenco1995elementary}. For this reason, and as commonly adopted in the literature, our discussion will center on this universal set of gates throughout the remainder of this paper. While our primary focus will be on the two-qubit CNOT gates, given their critical role in DQC, this emphasis does not limit the generality of our approach. Our model could also accommodate single-qubit gates by integrating the necessary parameters.

Figure~\ref{fig:quantum_circuit} depicts a quantum circuit comprising $7$ qubits and exclusively CNOT operations. A quantum circuit can also be described via a Directed Acyclic Graph (DAG), which captures the dependency relations between the gates. In this representation, the immediate set of executable gates, referred to as the \emph{frontier} is denoted as $F(G)$, where $G$ is the DAG.

Quantum entanglement is crucial for quantum communication and can be established between QPUs using quantum links, like photonic qubits through a fiber optic link. However, the process of generating entanglement is probabilistic, with success probabilities varying between $0$ and $1$, dependent on the hardware capabilities. The primary role of quantum network devices is to distribute EPR pairs across distant QPUs, overcoming the challenge of qubit fragility.

\begin{figure}
\centering
  \begin{subfigure}{0.7\columnwidth}
  \centering
 \includegraphics[width=0.8\textwidth]{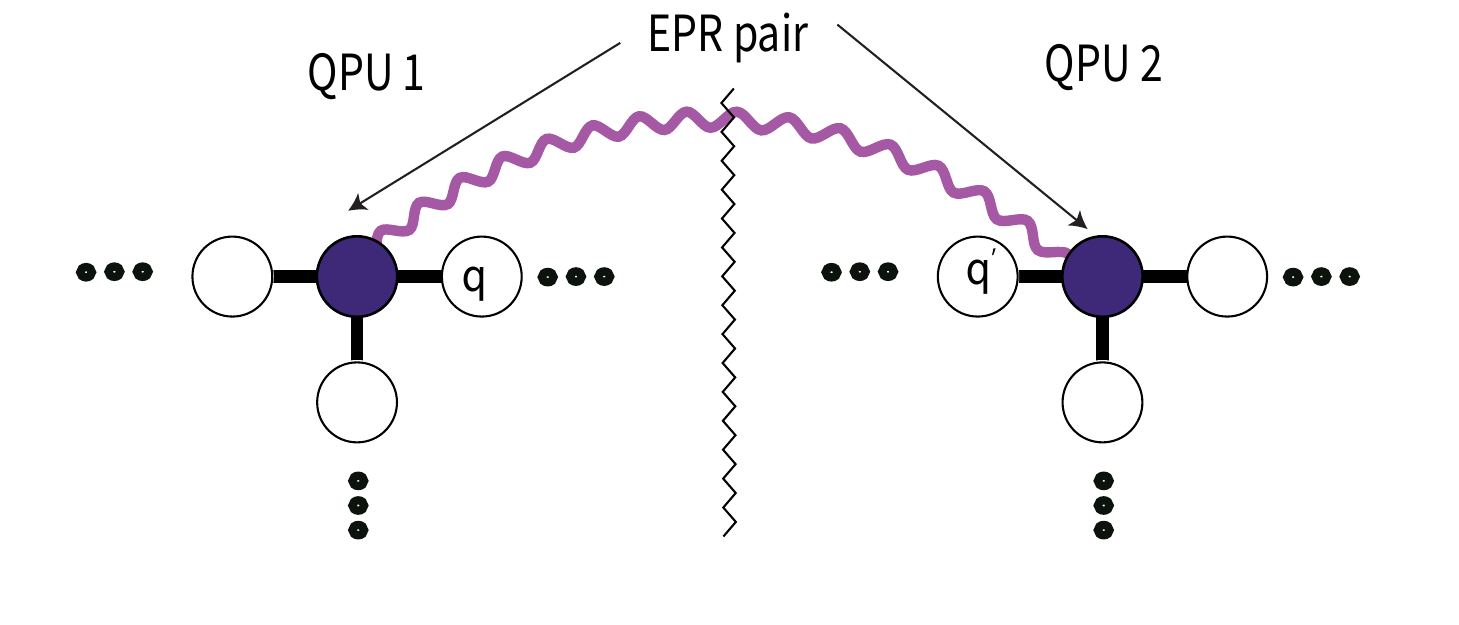}
  \vspace{-0.5cm}
  \caption{}
  \end{subfigure}
  \begin{subfigure}{0.7\columnwidth}
  \centering
  \includegraphics[width=0.8\textwidth]{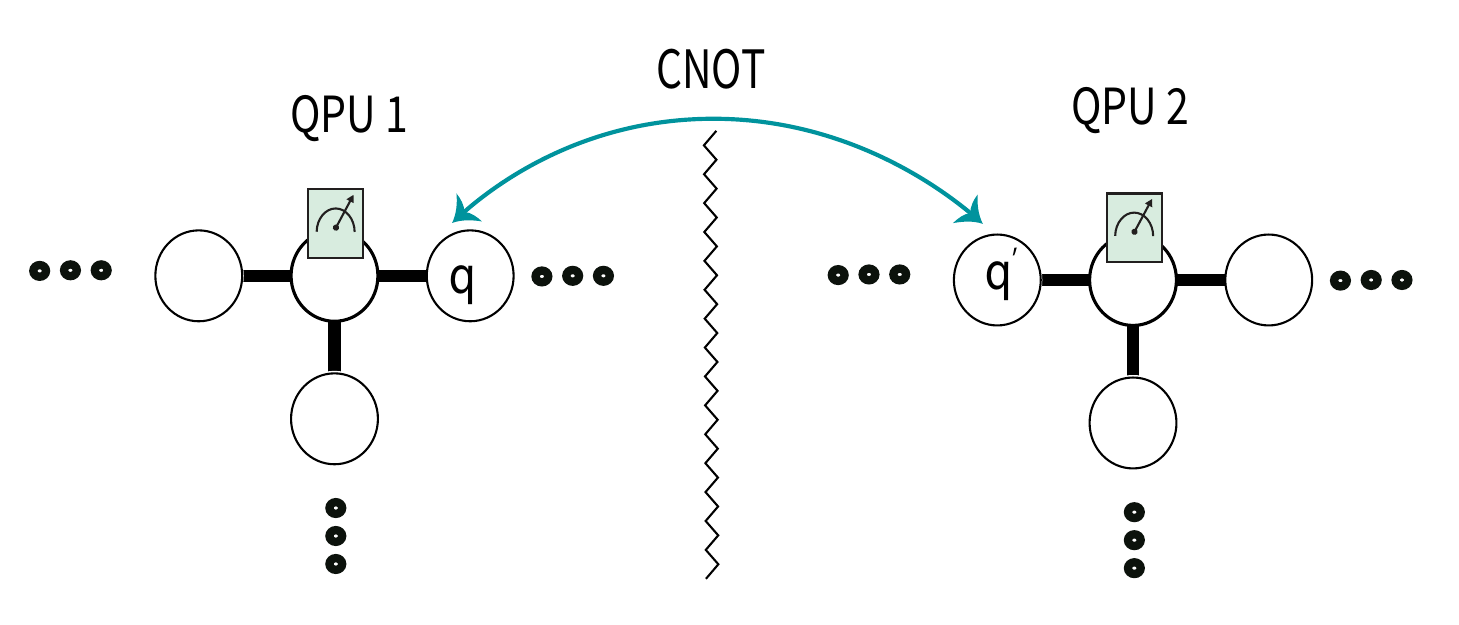}
  \vspace{-0.3cm}
  \caption{}
  \end{subfigure}
  \begin{subfigure}{0.7\columnwidth} 
  \centering
  \includegraphics[width=0.8\textwidth]{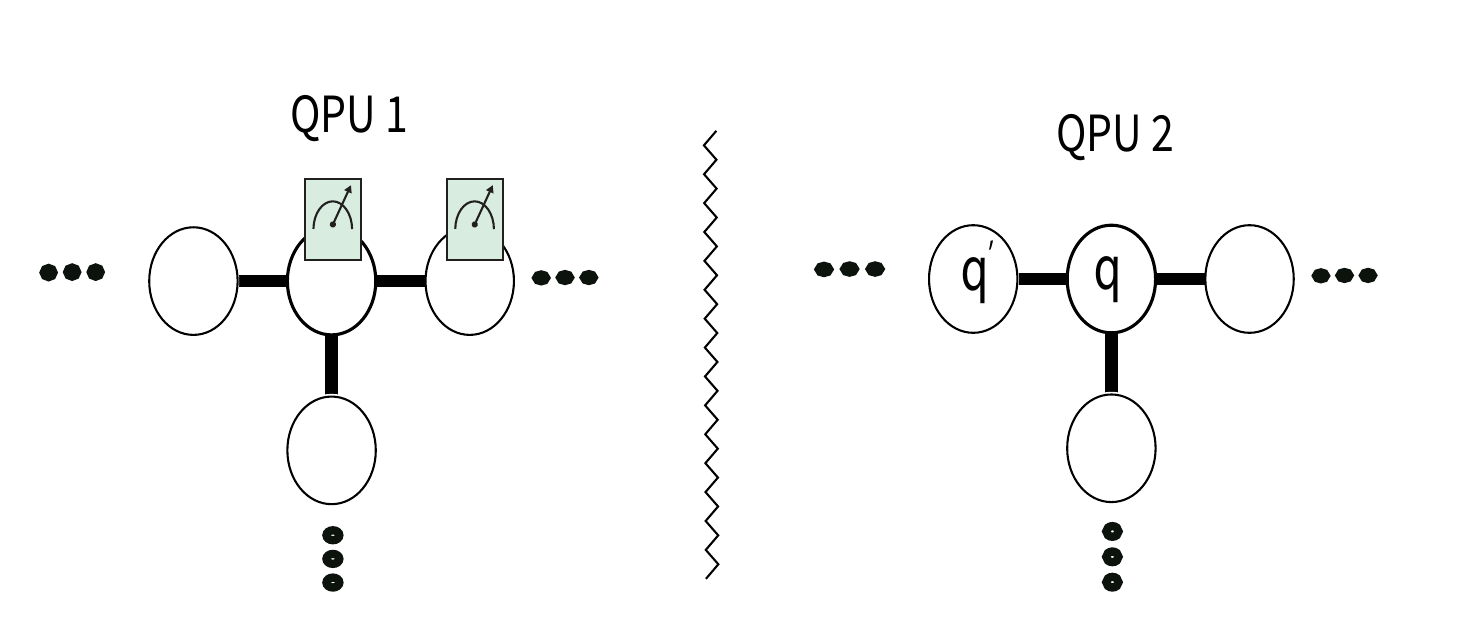} 
  \vspace{-0.3cm}
  \caption{} 
  \end{subfigure}  
  \vspace{-0.3cm}
  \caption{(a) Illustrates an EPR pair shared between two QPUs which can be used to teleport gates and qubits, (b) illustrates the state of the QPUs after a gate teleportation operation, while (c) shows the state of the QPUs after a qubit teleportation.}
  \label{fig:dqc_teleportations_operations}
  \end{figure}

The DQC environment design enables the implementation of quantum operations involving qubits across multiple QPUs. This is made possible by leveraging the EPR pairs, which can be utilized in two ways. Firstly, they enable the execution of a CNOT gate between physical qubits residing in different QPUs (gate teleportation). Secondly, they facilitate the physical teleportation of one qubit to the QPU where the other qubit is located, allowing for the local execution of the CNOT operation (qubit teleportation).

The outcomes of gate and qubit teleportations are depicted in Figure~\ref{fig:dqc_teleportations_operations}. In the case of gate teleportation (Figure~\ref{fig:dqc_teleportations_operations}(b)), the EPR pair is consumed to enable a remote CNOT operation between the logical qubits $q$ and $q'$. It is important to note that for this operation to occur, both qubits must be adjacent to the entangled qubits forming the EPR pair. On the other hand, in the context of qubit teleportation (Figure~\ref{fig:dqc_teleportations_operations}(c)), the EPR pair is used to transfer the state of one qubit (in the example qubit $q$) to the physical qubit that previously held the one EPR qubit. As a result, the EPR pair is destroyed.

\subsection{Quantum Processing Units Architecture}
\label{ssec:QPU_architecture}

Quantum computing technologies like superconducting circuits, ion traps, quantum dots, and neutral atoms show promise for advancing the field. Superconducting quantum circuits, in particular, are a focus for both academia and industry due to their potential. However, these technologies face limitations in architecture, notably in two-qubit gate implementation. Not all qubit pairs in a system can perform quantum operations directly due to physical constraints in hardware design. This leads to the concept of a quantum processor's \emph{coupling graph}, where nodes represent qubits and edges indicate possible two-qubit operations, focusing on CNOT operations for simplicity.

We formally denote the coupling graph of a QPU as a graph $P = (V, E)$ where $V$ represents the nodes that correspond the set of physical qubits and $E$ the set of links between the physical qubits. Let us also denote as $Q$ the set of logical qubits that are introduced by a quantum circuit. 

We define a qubit allocation as a function $f : Q \rightarrow V$ that maps logical qubits to physical qubits inside the QPU. Although the qubit allocation mapping is not an isomorphism due to possible empty qubit memories, we slightly abuse the notation by using $f^{-1}: V \rightarrow Q$ to denote what logical qubits are mapped to an actual physical qubit position inside the QPU. Finally if $e \in E$, where $e = (v_1, v_2), v_1,v_2 \in V$, we define $f^{-1}(e) := (f^{-1}(v_1), f^{-1}(v_2))$ to be the pair of logical qubits that are mapped to the physical qubits that the link $e$ connects. Since we consider our system to operate in a time slot manner and the logical qubits might change position in a given time slot, we have a different mapping $f_t$ for every time slot $t$. Therefore, $f_t(q)$ denotes the physical qubit that the logical qubit $q \in Q$ is mapped to at time slot $t$.

\begin{figure}[!t]
\centering
\includegraphics[width=0.3\textwidth]{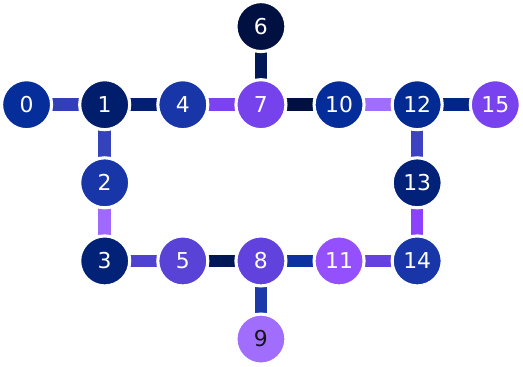}
\vspace{-0.25cm}
\caption{The coupling graph of the IBM Q Guadalupe quantum processor. This processor's type is Falcon r4P and can hold up to $16$ qubits. We refer interested readers to \cite{ibm2024quantum}, where IBM provides a list of the quantum processors and their corresponding coupling graphs.}
\label{fig:qpu_architecture_guadalupe_16_falcon}
\end{figure}

Figure~\ref{fig:qpu_architecture_guadalupe_16_falcon} depicts the coupling graph of the IBM Q Guadalupe quantum processor. The node color in Figure~\ref{fig:qpu_architecture_guadalupe_16_falcon} expresses the readout assignment error and the color in the links the CNOT error. Light color corresponds to greater error.

\subsection{Distributed Quantum Computing Architecture}
\label{ssec:DQC_architecture}


The establishment of EPR pairs and their distribution between QPUs is facilitated by quantum links. The physical qubits that are interconnected through the quantum links are referred to as \emph{link qubits}. We can swap the EPR pair halves to different physical qubit memories inside the QPU architecture. This capability allows for the pre-generation of EPR pairs, which can be strategically positioned adjacent to the qubits that require teleportation through SWAP operations. Additionally, by vacating a link qubit, new entanglement can be generated before the initial entanglement is utilized.


We denote as $\mathcal{C}$ the set of QPUs in the system and we use $C_i$ to denote the capacity in terms of physical qubits of QPU $i \in \mathcal{C}$. $Q$ represents again the logical qubits that the quantum circuit needs. However, in a time slot $t$ there might exist EPR pairs residing in physical qubits. Therefore, we extend the set of logical qubits from $Q$, to $\tilde{Q}_t$ to include the "alive'' EPR pairs in time slot $t$. We partition $\tilde{Q}_t$ into two disjoint set of logical qubits; $\tilde{Q}_t = Q + Q^{EPR}_t$, where $Q^{EPR}_t$ contains the qubits that are generated as parts of EPR pairs and exist in the QPUs at time slot $t$. Such qubits, wait to be consumed by the compiler for a teleportation operation. We introduce the mapping $\phi_t : Q_t^{EPR} \rightarrow Q_t^{EPR}$, that takes as an input a half of an EPR pair and gives as an output the other half that it is entangled to. For example, $\phi_t(q) \in Q^{EPR}_t$ and $q \in Q^{EPR}_t$ form an EPR pair, where the one half is located at $f_t(q)$ and the other at $f_t(\phi_t(q))$ (possibly in different QPUs). In a similar manner, in the rest of the paper we will use the tilde symbol, $\tilde{ }$, when we need to differentiate notation from Section~\ref{ssec:QPU_compiler} where maximally entangled pairs were not present.

We unify the coupling graphs of the QPUs by creating a single coupling graph $P = (V, E)$ as follows. We keep the physical qubits as vertices and we partition the edge set $E$ into two disjoint sets: $E = E^{p} + E^{n}$, where $E^{p} \cap E^{n} = \O.$ In set $E^{p}$ we keep the links that correspond to the couplings between physical qubits in individual QPUs whereas in $E^{n}$ we keep the quantum links that generate EPR pairs.

\section{Quantum compilers - Optimality Through an MDP}
\label{sec:quantum_compilers}

In this section we model the optimal compiler for a DQC environment using an MDP formulation. To achieve this goal, Section~\ref{ssec:QPU_qubit_mapping_and_compilation} discusses the problems of \emph{initial qubit mapping} and \emph{qubit routing}, with the latter being the primary focus of the compiler. Sections~\ref{ssec:QPU_compiler}~and~\ref{ssec:DQC_compiler} formulate the optimal compiler of a QPU and DQC respectively.

\subsection{Initial Qubit Mapping and Qubit Routing}
\label{ssec:QPU_qubit_mapping_and_compilation}

Quantum compilation translates the theoretical design of a quantum circuit, often created without accounting for specific hardware constraints, into a form that can be executed by a physical quantum computer. This process aims to preserve the intended quantum operations and outcomes while adhering to the architectural limitations of the target quantum hardware.

\begin{figure}[h]
\centering
\includegraphics[width=0.35\textwidth]{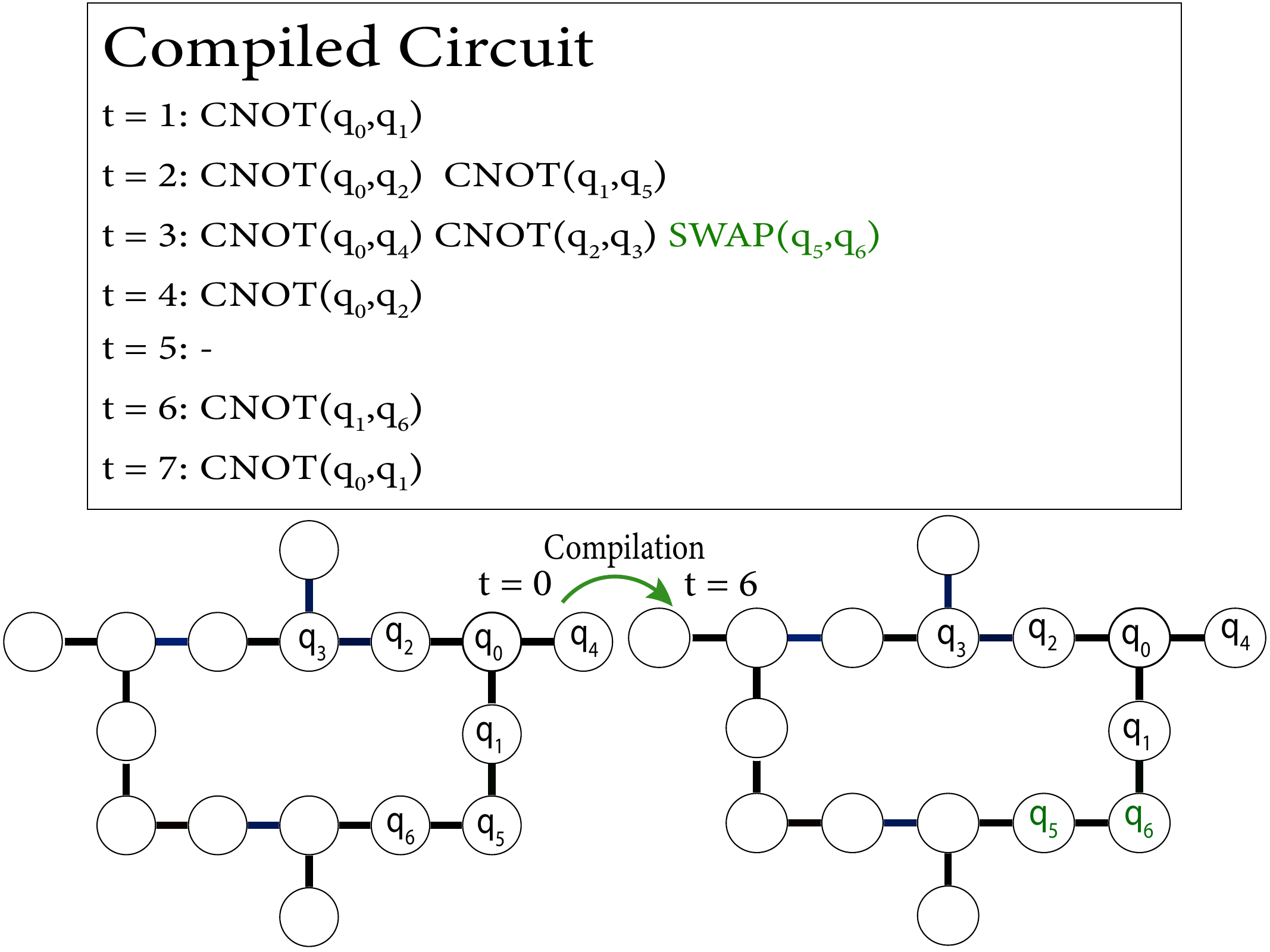}
\vspace{-0.3cm}
\caption{Illustration of one possible compilation of the circuit illustrated in Figure~\ref{fig:quantum_circuit} for IBM Q Guadalupe (see Figure~\ref{fig:qpu_architecture_guadalupe_16_falcon}).}
\label{fig:compiled_circuit}
\end{figure}

To exemplify the objective of a quantum compiler, let's direct our attention to the circuit represented by  Figure~\ref{fig:quantum_circuit}. For the purpose of this discussion, we will assume that the circuit needs to be compiled for the IBM Q Guadalupe architecture, depicted in Figure~\ref{fig:qpu_architecture_guadalupe_16_falcon}. Initially, it is worth noting that the circuit requires 7 qubits, indicating that our processor is equipped to successfully execute the computation. However, it becomes evident that there is no initial qubit placement that would allow the circuit's gates to be implemented without altering the logical qubit positions using SWAP operations.

The goal of the (single QPU) compiler is to introduce SWAP operations into the circuit, thereby enabling its execution while considering the constraints imposed by the hardware. Gates can be executed simultaneously in case they concern different qubits and thus the compilation should try to parallelize the operations as much as possible. In Figure~\ref{fig:quantum_circuit}, the first gate cannot be executed simultaneously with the second, but the second and third gates can be. Gates executable at the same time form a \emph{layer} of the circuit.
The resulting circuit, which incorporates the injected SWAP gates, is commonly referred to as the compiled circuit.
Figure~\ref{fig:compiled_circuit} illustrates one possible compilation of the circuit illustrated in Figure~\ref{fig:quantum_circuit} for IBM Q Guadalupe. Given that the initial qubit allocation is as shown in the $t=0$ of the figure, it illustrates a possible compiled quantum program and the state of the quantum processor unit before and after the injected SWAP gate. We define a time slot as the time needed for the QPU to execute a CNOT gate.

In DQC, the compiler should also inject remote operations (e.g., qubit and gate teleportations) while also requesting entanglement generations. Specifically, for a quantum circuit to be effectively executed, two distinct procedures are required:

\textbf{Initial qubit mapping:} Firstly, the logical qubits of the circuit should be mapped to the physical qubit memories of the various QPUs. This can be seen as the starting point of the compilation and thus is very important for its efficiency. In Figure~\ref{fig:compiled_circuit} observe that even in the single QPU case different initial qubit mapping would possibly need more SWAP operations in the compiled circuit (resulting in increased latency).

\textbf{Qubit Routing/Quantum Compilation:} After the compiler gets an initial qubit mapping, it should (i) generate and route the EPR pairs, (ii) schedule remote operations, and, (iii) inject SWAP gates to facilitate the execution of local gates inside the QPUs. Efficiently finding the optimal compilation is crucial in quantum computing, as quantum resources are scarce and fragile. The compiled circuit exhibits differences in both the number of layers and the number of gates compared to the initial quantum circuit. Note that the number of layers in the initial circuit is not indicative of the total running time of the quantum circuit since there are injected gates that should be implemented in the compiled version. This fact illustrates why the initial qubit mapping should not be based only on the initial circuit but rather should be jointly optimized considering the compiler model (see discussion in Section~\ref{sec:discussion_and_model_extension}).



\subsection{Optimal Compiler for a Single QPU: MDP Formulation}
\label{ssec:QPU_compiler}

In this section, we will develop a model for the QPU compiler with the objective of minimizing the expected execution time of a quantum circuit. To accomplish this, we will define (i) the state space of the QPU, which represents its current configuration, and (ii) the action space, from which the compiler selects actions in each time slot. Using a dynamic framework, the compiler will make optimal decisions, time slot by time slot, to construct the compiled circuit.

\textbf{State Space:} Recall that $P = (V,E)$ denotes the coupling graph of the QPU under consideration, where $V$ denotes the nodes that represent the set of physical qubits and $E$ the set of links between the physical qubits. Let a physical qubit $v \in V$, we use $\delta_V(v)$ to denote $v$'s neighboring qubits and $\delta_E(v)$ to denote the links that emanate from $v$.

Note that the execution of a SWAP gate requires three CNOT gates. Consequently, when a SWAP gate is applied to a link, the link becomes temporarily unavailable until the completion of the gates. Essentially, if a SWAP gate involves a physical qubit $v \in V$, then all the links in $\delta_E(v)$ are blocked until the gate is finished. To capture the availability of links within a specific time slot $t$, we introduce the function $c_t : V \rightarrow \mathbb{N_+}$ that denotes the number of time slots that must elapse before a physical qubit can be used again.



For a given time slot $t$, the state of the system is defined as $s_t = (f_t, c_t,G_t) \in \mathcal{S}$, where $f_t$ describes the placement of logic qubits within the QPU, $c_t$ denotes the nodes' cooldowns, i.e., the time needed until the node/qubit can be used again, and $G_t$ represents the remaining DAG that consists of the operations that are to be executed for the completion of the initial circuit. Note that the feasible gate operations are dictated by the frontier of $G_t$, $F(G_t)$. The ultimate objective of the compiler is to reach a state $s^{*}_{t^*} = (\cdot, \cdot, \O)$ at a (preferably minimal) time slot $t^*$, where $\O$ denotes an empty graph, indicating the successful completion of the compilation.



\textbf{Action Space:} The action space captures all the possible actions of the compiler. For every link in the coupling graph, the compiler can perform one of the following actions: (i) nothing (from now on denoted as $\O$), (ii) CNOT gate, or (iii) SWAP gate. The compiler can execute multiple CNOT gates to neighboring physical qubits at a time slot. Therefore, the actions of the compiler correspond to \emph{matchings}\footnote{A matching is a subset of edges in a graph such that no two edges share a common vertex.} in the coupling graph, where each link $(u,v)$ of the matching corresponds to a gate executed on physical qubits $u$ and $v$.

\underline{Actions correspond to matchings;} Let us denote the set of matchings of a coupling graph $P$ as $\mathcal{M}(P)$. $m \in \mathcal{M}(P)$ represents a matching of the coupling graph, which consists of edges that link physical qubits. However, a link $e = (u,v) \in E$ of the coupling graph $P$ cannot be used in a time slot in case either $c_t(u) > 0$ or $c_t(v) > 0$ by definition. Therefore, to indicate the links that we can activate in every time slot we can construct a new, truncated graph  denoted as $P^{tr}_t = (V_t^{tr}, E^{tr}_t)$, where the set $V^{tr}_t \subseteq V$ does not include nodes that have cooldowns associated to them. Therefore, 
\begin{align*}
 V^{tr}_t := \Big\{ v\in V  : c_t\left(v\right) >0  \Big\}.
\end{align*}
Similarly, $E_t^{tr}$ includes the links from $E$ that connect nodes in $V_t^{tr}$, i.e, $E_t^{tr} := \left\{ (u,v) \in E : u,v \in V_t^{tr}  \right\} $. Hence, to express the set of links that we can enable in every time slot we focus on the set of matchings of $P_t^{tr}$, i.e., $\mathcal{M}\left(P_t^{tr}\right)$. 
We map the actions to matchings in this graph and we assume that the non existence of a link in a matching means that for this link we pick the null action, $\O$.

\underline{Actions are matchings with (unique) labels on the links;} However, the activation of a link (or existence of a link in a matching) can represent two different operations, (i) a CNOT execution, or (ii) a SWAP gate. To formalize the action set, we introduce a labeling mapping of the edges of the graph $P$ at time slot $t$, and denote it as $l_t : E \rightarrow \{\text{"swap", "score"}\}$. The compiler should pick a labeling mapping in every time slot $t$ according to which ''SWAP" in a link $e \in E$ would mean that a $SWAP(f^{-1}(e))$ would be injected at time slot $t$, and the label "score" in a link $e \in E$ would mean that a $CNOT(f^{-1}(e))$ would be executed in that time slot. However, depending on the state $s_t$ of the system not all possible labels are feasible for a link.

The "score" label can be put in a link $e \in E_t^{tr}$ only in case $CNOT(f^{-1}(e)) \in F(G_t)$. On the contrary the "swap" label can always be used. However, a little thought reveals that there is not any rationale for the optimal compiler to inject a "swap" before it injects a "score" to a pair of qubits in case that is possible. We introduce the set of eligible labelings for a matching $m \in \mathcal{M}(P_t^{tr})$ at time slot $t$, given the state $s_t$ as follows:
\begin{align*}
& L^{el}(s_t,m) := \Big\{  l: l(e) = \O, \forall e \notin m, \\
& \qquad l(e) = "swap", \forall e \in m: CNOT(f^{-1}(e)) \notin F(G_t), \\ 
& \qquad l(e) = "score", \text{o.w} \Big\}
\end{align*}
Observe that there is a unique label for every activated link of the matching and thus we can rigorously define the set of actions of the compiler given a state $s_t$ at time slot $t$ as:

\begin{equation*}
\mathcal{A}(s_t) = \Big\{ m : m \in \mathcal{M}(P_t^{tr})  \Big\}. 
\end{equation*}
\textbf{System Evolution:} After picking a strategy $a_t \in \mathcal{A}(s_t)$ in a time slot $t$, the system state evolves to the next state, $s_{t+1}$, according to the following intuitive recipes:
\begin{itemize}
    \item For $f_{t+1}$ the only elements that change correspond to the qubits for which we implemented a SWAP gate. Specifically, $f_{t+1}(q) = f_t(q^{'}),  f_{t+1}(q) = f_t(q^{'})$ for every $SWAP(q,q^{'})$ which was injected at time slot $t$.
    \item For $c_{t+1}$ first we subtract one time slot for every non positive weight, i.e., $c_{t+1}(u) = c_t(u) - 1, \forall u:c_t(u)>0$. Secondly, for every injected $SWAP(f^{-1}_t(u,v))$ at time slot $t$, $c_{t+1}(u) = c_{t+1}(v) = 2$ (equivalent to 3 CNOT). 
    \item For $G_{t+1}$ we should remove the gates that were implemented with the label "score" at time $t$.  
\end{itemize}

Having formulated the optimal compiler for a single QPU as an MDP, we express its task through a cost function quantifying the time required to complete the DAG. This conversion into an optimization problem is deferred to Section~\ref{ssec:DQC_compiler}, where we immediately address the optimization challenge for the optimal compiler in a DQC setting.

\subsection{Optimal Compiler for DQC: MDP Formulation}
\label{ssec:DQC_compiler}

In this section, we will build upon Section~\ref{ssec:QPU_compiler} by extending the MDP formulation to develop an optimal compiler for a DQC environment featuring multiple interconnected QPUs.

To execute a quantum circuit, the logical qubits should be placed to physical qubits, now potentially in different QPUs. Subsequently, the compiler manages qubit routing. This process involves generating and handling EPR pairs, scheduling remote operations, and injecting SWAP gates to facilitate the execution of local gates. Although in the rest of the section we model the compiler through its optimal qubit routing, in Section~\ref{sec:discussion_and_model_extension} we discuss how we can use the compiler to find an initial qubit mapping.

\textbf{State Space:}  
As in Section~\ref{ssec:QPU_compiler}, we use $f_t$ to denote the qubit mapping, $c_t$ to express the cooldown weights for every node $v \in V$, and $G_t$ to denote the DAG to be executed for a time slot $t$. We redefine the state of the system at time slot $t$ as a vector $s_t = (Q^{EPR}_t, f_t, c_t, G_t)$, where $Q^{EPR}_t$ denotes the  EPR pairs available at time $t$. $Q^{EPR}_t$ is assumed to include the mapping $\phi_t$ (see Section~\ref{ssec:DQC_architecture}) necessary to spot the entangled halves of the EPR pairs. The ultimate objective of the compiler again is to reach a state $s^*_{t^*} = (\cdot, \cdot, \cdot, \O)$ at a (preferably minimal) time slot $t^*$.


\textbf{Action Space:}
The action space is now enriched with operations that concern the DQC framework. Specifically, we now have the following operations/actions (i) nothing (denoted as $\O$), (ii) SWAP gate,  (iii) CNOT gate, (iv) teleport gate, (v) teleport qubit, and (vi) EPR generation. Once again, a constraint imposed by the hardware is that when a qubit is involved in an operation, no other gate can affect its state. Therefore, to introduce the action space, we should focus again on the matchings of a graph. Nevertheless, the teleportations, which involve qubits entangled among different QPUs makes the coupling graph and the trancated graph $P, P_t^{tr}$, developed in the Section~\ref{ssec:QPU_compiler} no more useful.

For that reason, we introduce a hypergraph\footnote{A hypergraph is a generalization of a graph in which an edge can join any number of vertices. In an ordinary graph, an edge connects exactly two vertices.} $\tilde{P}_t = (V,\tilde{E}_t)$ that includes also hyperlinks that capture the teleportation operations. We create the set $\tilde{E}_t$ of the hyperlinks by partitioning it into $3$ disjoint sets, i.e.,  $\tilde{E}_t = E + E^{tg}_t + E^{tq}_t$, where $E = E^{p} + E^{n}$ is the set of edges of the graph $P$ as usual,
\begin{align*}
& \tilde{E}^{tg}_t := \Big\{ (u, f_t(q), f_t(\phi(q)), v),  \forall q \in Q^{EPR}_t, \\ & \qquad \forall u \in \delta_V(f_t(q)), \forall v \in \delta_V(f_t(\phi(q))) \Big\} \text{, and,}
\end{align*}
\begin{align*}
& \tilde{E}^{tq}(t) := \Big\{ (v, f_t(q), f_t(\phi(q))),  \forall q \in Q^{EPR}_t, \\ & \qquad \forall v \in \delta_V(f_t(q)) \Big\}.
\end{align*}
The set $E^{tg}$ encompasses hyperlinks signifying the possible teleportation of gates, while the set $E^{tq}$ contains hyperlinks that denote the teleportation of qubits. It should be noted that when a hyperlink from either set is activated, the nodes involved in the operation must be rendered inactive until the operation concludes.

\underline{Actions correspond to matchings;} Similarly with Section~\ref{ssec:QPU_compiler}, we construct the truncated graph of $\tilde{P}_t$ for a time slot $t$, $\tilde{P}^{tr}_t = (V^{tr}_t, \tilde{E}^{tr}_t)$, by excluding nodes that have cooldowns. Therefore, 
\begin{align*}
 V^{tr}_t := \Big\{ v: v\in V  \text{ and } c_t(v) >0 \Big\}.   
\end{align*}

$\tilde{E}^{tr}_t$ includes the links from $\tilde{E}_t$ that connect nodes in $V^{tr}_t$.
    Hence, to express the set of links that we can activate in every time slot we focus on the set of matchings\footnote{A matching $m$ on a hypergraph is a set of hyperedges such that every two hyperedges in $m$ have an empty intersection (have no vertex in common).} of $\tilde{P}^{tr}_t$, i.e., $\mathcal{M}(\tilde{P}^{tr}_t)$. Observe that a matching $m \in \mathcal{M}(\tilde{P}^{tr}_t)$ contains links that can be activated simultaneously in a time slot. However, the actual action that will be associated with a link is to be considered using labels as in Section~\ref{ssec:QPU_compiler}.

\underline{Actions are matchings with (unique) labels on the links;} To formalize the action set, we introduce a labeling mapping of the edges of the graph $\tilde{P}_t$ at time slot $t$, and denote it as $l_t : \tilde{E}_t \rightarrow \{\text{\footnotesize "swap", "score", "tele-gate","tele-qubit", "generate"}\}$. The compiler picks a matching with labeled edges in every $t$ according to which "swap" and "score" in a link injects a SWAP and a CNOT gate respectively. "tele-gate" in a (hyper)link $e = (v_1,v_2,v_3,v_4) \in E^{tg}_t$ injects a $CNOT(f^{-1}( (v_1,v_4) ))$ (see Figure~\ref{fig:dqc_teleportations_operations}(b)). Labels "tele-qubit" and "generate" will not affect the compiled circuit explicitly but only the state transitions (see System Evolution below).



Similarly to Section~\ref{ssec:QPU_compiler}, we  introduce the set of eligible labelings for a matching $m \in \mathcal{M}(\tilde{P}^{tr}_t)$ at time slot $t$, given the state $s_t$ as follows:
\begin{align*}
& L^{el}(s_t,m) := \Big\{  l: l(e) = \O, \forall e \notin  m, \\
&  l(e) = "swap" , \forall e \in E^p \cap m: CNOT(f^{-1}(e)) \notin F(G_t), \\
&  l(e) = "score", \forall e \in E^p \cap m: CNOT(f^{-1}(e)) \in F(G_t) , \\
&  l(e) = "tele-gate" ,  \forall e = (v_1,v_2,v_3,v_4) \in E^{tg}_t \cap m: \\  & \qquad \quad  CNOT(f^{-1}_t(v_1,v_4)) \in F(G_t), \\
&  l(e) = "tele-qubit" , \forall e \in E^{tq}_t \cap m, \\
&  l(e) = "generate" , \forall e \in E^n \cap m: f_t^{-1}(e) = \O \Big\}.
\end{align*}
$L^{el}(s_t,m)$ assigns labels only to links with feasible operations available. For example, a link in $E^n$ can generate an EPR pair only if the link qubits associated are empty. Note that each activated hyperlink may be associated with a unique label, allowing us to characterize the action space by identifying matchings within the truncated graph. 
We can now define the compiler's action set for a given state $s_t$ at time slot $t$ as:
\begin{equation*}
\mathcal{A}(s_t) = \Big\{ m: m \in \mathcal{M}(\tilde{P}^{tr}_t ) \Big\}. 
\end{equation*}
\textbf{System Evolution:} After picking a strategy $a_t \in \mathcal{A}(s_t)$ in a time slot $t$, the system state evolves to $s_{t+1}$, as follows:
\begin{itemize}
    \item $Q^{EPR}_{t+1}$ (i) includes the new EPR generated from successful "generate" activations, and (ii) deletes the EPR pairs destroyed from "tele-gate" and "tele-qubit" operations.
     \item The mapping $f_{t+1}$ evolves from $f_{t}$, with modifications in certain elements, as detailed below ("swap" is omitted since its effect on $f_t$ is described in Section~\ref{ssec:QPU_compiler}):
     \begin{enumerate}
    \item The "tele-gate" operation in a hyperedge $(v_1,v_2,v_3,v_4)$ change $f_{t+1}$ as: 
    \begin{align*}
        &  f^{-1}_{t+1}(v_2) = f^{-1}_{t+1}(v_3) = \O. 
    \end{align*}
    \item The "tele-qubit" operation in a hyperedge $(v_1,v_2,v_3)$ change $f_{t+1}$ as: 
    \begin{align*}
        & f^{-1}_{t+1}(v_2) = f^{-1}_{t}(v_1),  \quad
          f^{-1}_{t+1}(v_1) = f^{-1}_{t+1}(v_3) = \O. 
    \end{align*}
    \item The "generate" operation in a link $e = (v_1,v_2) \in E^{n}$ - if successful - creates an EPR pair $q, q^{'} \in Q^{EPR}_{t+1}$ and changes $f_{t+1}$ as: 
    \begin{align*}
        & f_{t+1}(q) = v_1, f_{t+1}(q^{'}) = v_2.  
    \end{align*}
     \end{enumerate}
    
   \item For $c_{t+1}$ we should first subtract one for every non positive weight, i.e., $c_{t+1}(v) = c_t(v) - 1, \forall v:c_t(v)>0$. Secondly, we should update the weights for the nodes depending on the actions taken at time slot $t$. Note that both gate and qubit teleportation circuits can be executed in $\kappa + 1$ time slots \cite{nielsen2010quantum}, where $\kappa$ denotes the time needed for a classical bit to get transferred through classical links. The effect of "swap" is described in Section~\ref{ssec:QPU_compiler} and thus is omitted.

         \begin{enumerate}
         
    \item For every "tele-gate" operation in a hyperedge $(v_1,v_2,v_3,v_4)$: 
    \begin{align*}
        c_{t+1}(v_1) = c_{t+1}(v_2) = c_{t+1}(v_3) = c_{t+1}(v_4) = \kappa. 
    \end{align*}
     \item For every "tele-qubit" operation in a hyperedge $(v_1,v_2,v_3)$: $c_{t+1}(v_1) = c_{t+1}(v_2) = c_{t+1}(v_3) = \kappa. $
    \item For every "generate" operation in a link $e = (v_1,v_2) \in E^{n}$: $c_{t+1}(v_1) = c_{t+1}(v_2) = \lambda - 1,$
    
     \end{enumerate}
where $\lambda$ (potentially $\neq \kappa$) denotes the time needed for the entanglement generation protocol to finish.
    \item For $G_{t+1}$ we should remove nodes from $G_t$ according to (i) the CNOT gates, and, (ii) the gate teleportations we implemented at time slot $t$. In the case of a "tele-gate" operation, it is important to note that only after $\kappa$ time slots, when then the action will be completed, we will be able to delete the corresponding node of the DAG.

\end{itemize}

\textbf{Objective: } The mission of the compiler is to produce a circuit with the minimal (expected) feasible depth, optimizing both its execution speed and the likelihood of successful operation by the QPU. Let $N$ represent a time horizon, expressed in the number of time slots, within which we are certain that qubit decoherence will occur. Therefore, the objective is to execute the circuit as rapidly as possible\footnote{If the probability for successful entanglement is one, the time elapsed until the compilation of the circuit coincides with the compiled circuit's depth.}, and definitively before reaching $t = N$. Let $g(s_t)$ denote the cost incurred at time slot $t$ for a state $s_t$. We encode the need for minimizing the expected execution time of the circuit into $g$:
\[ 
g(s_t) = \left\{
\begin{array}{ll}
      \infty & t = N, G_t \neq \O \\
      0 &  G_t = \O \\
      1 & \text{otherwise.}  \\
\end{array} 
\right. 
\]
Because of the intrinsic stochasticity of the framework, the objective of the compiler is to pick actions/matchings $a_1, \dots, a_N$ to minimize the \emph{expected} cost:
\begin{equation}
\label{eq:objective_of_MDP}
    min_{a_1, \dots, a_N} \ \mathbb{E} \Big\{ g(s_N) + \sum_{t=0}^{N - 1} g(s_t) \Big\}.
\end{equation}

Section~\ref{sec:quantum_compilers} presented a model for the optimal compiler. Nevertheless, the vast array of potential states and actions within the system makes analytical solutions via MDP formulation potentially infeasible. Nonetheless, in Section~\ref{sec:reinforcement_learning_implementation}, we modify this model, enabling the derivation of approximately optimal compilers for the DQC framework.

\section{Reinforcement Learning Implementation}
\label{sec:reinforcement_learning_implementation}

Reinforcement Learning has been widely used as an efficient method for deriving  policies for MDPs, especially when the environment under which the policy needs to be developed is sufficiently complicated. There are many model-free off-policy (e.g. Deep Q-Network (DQN) \cite{mnih2013playing}, Double Deep Q-Network (DDQN) \cite{van2016deep}) or on-policy (policy gradient \cite{van2016deep}, Proximal Policy Optimization (PPO) \cite{schulman2017proximal}) RL methods being used or improved upon, which have led to remarkable achievements in various fields \cite{silver2018general, kober2013reinforcement}. The MDP formulation in Section~\ref{ssec:DQC_compiler} happens atop a highly complex environment; a processor capable of holding $|V|$ different qubits will have, depending on the DAG required to be solved, $|\tilde{Q}_t|$ different logical qubits present in the processor. Furthermore, each action (from the possible $|\tilde{E}_t|$ actions) could have a cooldown period of up to $C_{max}$.\footnote{$C_{max}$ denotes the maximum cooldown possible.} This alone would define a state space of size $(|V|!/(|V|-|\tilde{Q}_t|)!)(|\tilde{E}_t|*C_{max})$, leading to a very large state space that would be impractical even for traditional RL methods like Q learning. As a result, we introduce neural network-driven learning agents such as DQN, DDQN and PPO, as a more feasible way to train the learning agent (approximators), which are known to be efficient at finding policies for MDPs with large state and/or action space \cite{mudvari2023robust}.

\subsection{Efficient formulation and constrained RL}

It is well-documented that RL techniques begin to suffer as the dimensions of the state and the action space grow \cite{dulac2019challenges}. So a sound strategy would be to consider different methods that would allow for the problem to be represented in a way that reduces these dimensions. In the following, we describe the state and action spaces used in our RL formulation, highlighting how we simplified or approximated them compared to the optimal formulation discussed in Section~\ref{ssec:DQC_compiler}.

{\bf State Space:} The (reduced) state space considered for the RL formulation\footnote{The state space of the \textit{actual} environment of the DQC framework will be exactly as formulated in Section~\ref{ssec:DQC_compiler}. However, we should reduce the information that we feed the NN agent for efficiency.} consists of three components, with the first two components serving as input to the NN learning agent. The first component specifies the location of the logical qubits on the processor, represented as a vector of size $|V|$, denoted as $S_{loc}$.\footnote{$S_{loc}$ corresponds to the mapping $f^{-1}$ introduced in Section~\ref{ssec:QPU_architecture}.}  
The second component of the state represents the DAG being solved, and is a vector $S_{dag}$ of size $3G_{max}$. $G_{max}$ denotes the number of gates in the circuit, and the multiplication factor of 3 is there because each DAG vertex is identified with 3 integers: the first two elements are the logical qubits that need to undergo "score" operation, and the third component denotes the layer of the corresponding gate in the DAG.
The third and final component of the state vector, $S_{msk}$, is a mask vector with a dimension equal to the number of considered actions (described below), indicating the availability of each action depending on the state $s_t$.  Thus the state vector is $S=[S_{loc},S_{dag},S_{msk}]$. 

\begin{approximation}
Note that $S_{msk}$ is a binary vector. Compared to the formulation in Section~\ref{ssec:DQC_compiler}, we have omitted the actual cooldown time for each link from the state. By doing this, depending on $C_{max}$, we significantly reduced the state space. The mask only indicates whether a label is available, without specifying the time required for it to become available again. The rationale behind this approximation is that the compiler will not benefit significantly from knowing \textit{exactly} when each link will be available again.
\end{approximation}

{\bf Action Space:} Recall from Section~\ref{ssec:DQC_compiler} that the action space of the optimal compiler consists of the set of matchings of the hypergraph $\tilde{P}_t$. It is only by defining the optimal action space as this set of matchings that we could formulate the problem as described in Eq.~\ref{eq:objective_of_MDP}. However, this action space is too large for an RL method to handle efficiently.

\begin{approximation}
We reformulate the action space to consist of \textit{all} feasible labels (as defined in Section~\ref{ssec:DQC_compiler}) for \textit{every} link. In other words, instead of matchings, we now enumerate all possible labels for the hypergraph’s links. Consequently, we introduce the "stop" action as an auxiliary action, corresponding to the end of a matching, which would reduce the cooldowns in the current state of the environment. Note that this approximation decouples an iteration of the RL agent from the actual time slot in the DQC environment's compilation process. Although we sacrifice optimality, through a well-designed reward shaping, we can align the RL agent's rewards with the actual time elapsed in the real DQC system.
\end{approximation}

\begin{approximation}
    Inspired by the observation that a "score" action\footnote{In the RL formulation \emph{labels} from Section~\ref{ssec:DQC_compiler} became \emph{actions}.} can always be executed immediately when possible without sacrificing optimality, we apply the same principle to the "tele-gate". Thus, whenever an EPR pair neighbors two logical qubits in a gate at the frontier of the DAG, we immediately teleport the gate.
\end{approximation}

Therefore, an action in the RL formulation can be either "stop", "swap" in any link of graph $P$, any possible "tele-qubit" operation, and finally "generate" for every link in $E^n$. Therefore, the action space has dimension $1 + 2  | E^p | + |E ^n|$. Note, that we do not define the subset of actions at each step depending on their feasibility, but we rather use $S_{msk}$ to help the learning agent only choose the actions that are feasible. Since the learning agent is NN-based, this requirement ensures that all actions can be assigned to the output layer of the NN.       

{\bf Rewards:} Depending on the state every action introduces a reward to the RL agent. Specifically, every time a "score" is implemented successfully, independently of the state, the agent gets $R_{score}$. Moreover, since we allow $N$ time slots (and thus $N$ "stop" actions) for the compiler to execute the circuit, we enforce a $R_{success}$ reward whenever it finished the circuit and a negative $R_{fail}$ if it fails. To map the timing of the actual DQC environment with the time slots of the RL agent we also incur a negative reward $R_{stop}$ every time the action is "stop".

For the "swap" and "tele-qubit" actions we get inspiration from \cite{pozzi2022using} to design a distance based reward shaping. We construct a graph, similar to $P$ but we put weight $1$ for every link in $E^{p}$ and a large weight, $w_{qlink}$, for every link in $E^n$. This captures that we need one SWAP to traverse a link inside a QPU but it is not possible for a qubit to change QPU without using an EPR pair. For that reason, we also connect the entangled EPR pairs with a link of weight $1$. We can now calculate a metric, $d(q,q^{'})$ that calculates the shortest distance between the logical qubits $q$ and $q^{'}$ in the aforementioned graph. To calculate the reward we use one more distance metric called $d^{frontier}_{t}$ that calculates and adds up the distances $d(q,q^{'})$ for \emph{every} gate $CNOT(q,q^{'})$ in the frontier. Every time in the aforementioned calculation that an EPR is used, we delete their corresponding link in the graph so that we do not use twice the same EPR for the calculation of $d^{frontier}_{t}$. For every time slot $t$, we calculate the reward of "tele-qubit"s and "swap"s with the reward $R_{dist} = \xi (d^{frontier}_{t-1} - d^{frontier}_{t})$, where $\xi$ is a parameter.

\subsection{Learning Agent and RL approaches}
\label{ssec:learning_agent}


\begin{figure}[h!]
\includegraphics[width=0.49\textwidth]{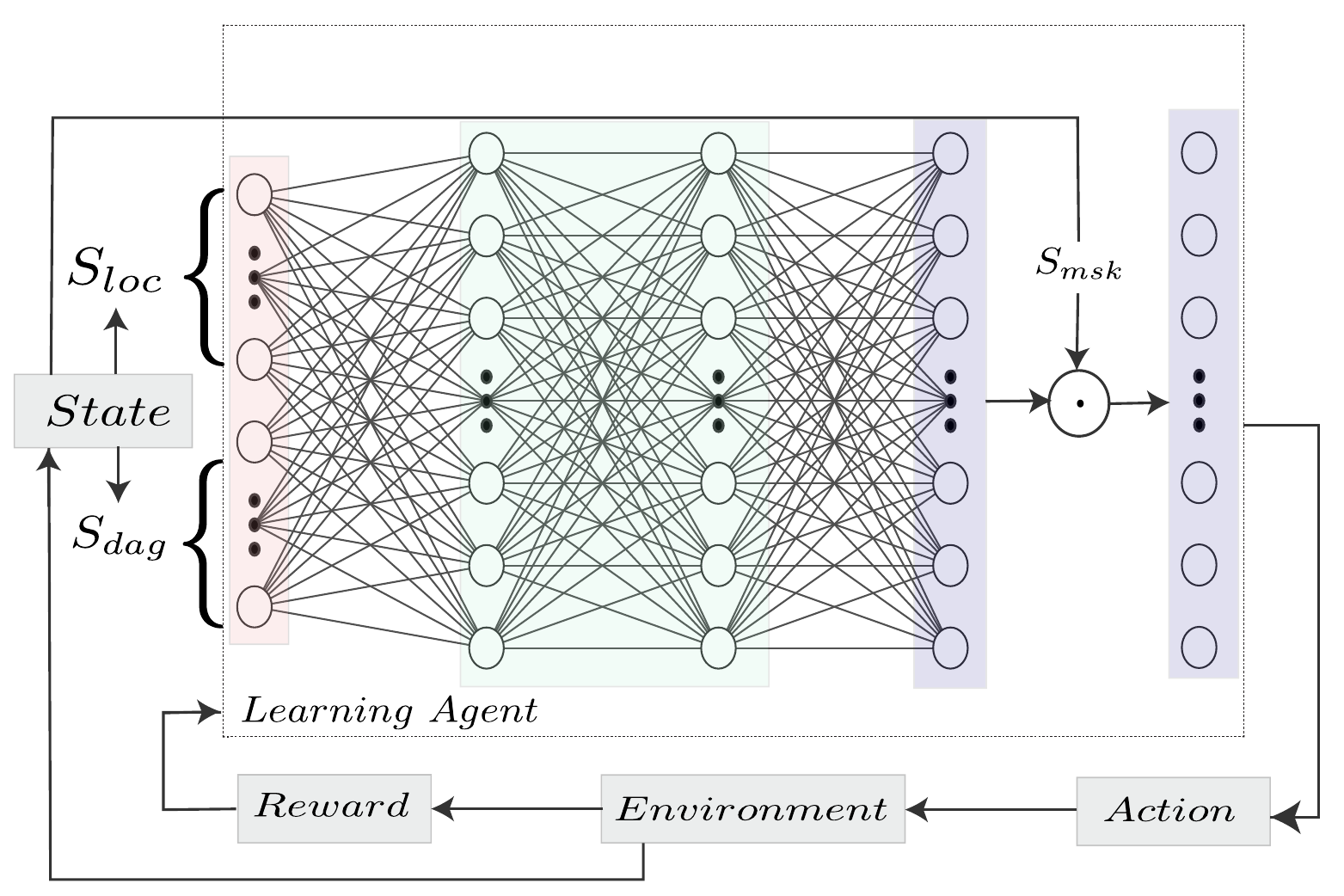}
\caption{Overview of the learning Agent (DDRL-based example) in the constrained RL environment.}
\label{fig:learning_agent}
\end{figure}

Figure~\ref{fig:learning_agent} illustrates the learning agent under the proposed RL environment, primarily referring to DQN/DDQN approach. For policy-based approaches, such as PPO, an additional step is required to ensure that the probabilistic selection of action remains within the constraints defined by $S_{msk}$. The state of the environment is accessed by the learning agent NN, where the first layer will take as input $S_{loc}$ and $S_{dag}$, while $S_{msk}$ is entered at second-last layer, where it ensures that infeasible actions always have a lower value than all the feasible actions (through Hadamard product of $S_{msk}$ and the second-last layer). For the value based methods, these values represents the Q-value of each actions given the state, and are outputs in last layer. In policy gradient methods, it's crucial to calculate the probabilities assigned to each action such that infeasible actions are assigned a probability of zero \cite{huang2020closer}, effectively enforcing $S_{msk}$ constraints.

\section{Discussion: Model Extensions}
\label{sec:discussion_and_model_extension}

In this section, we discuss potential extensions and applications for the model developed for our compiler. These ideas are based on initial concepts and will be explored through comprehensive simulation in our future work.
\vspace{0.1cm}

\textbf{Initial qubit mapping:} Our optimal compiler can identify an effective initial qubit mapping that aligns well with the compiled circuit. The process unfolds over three key steps (see Figure 5 in \cite{li2019tackling}): initially, the compiler compiles the original circuit, using the final qubit placements to establish the initial mapping for a compilation of the reverse circuit. It then compiles this reverse circuit. After executing the reverse circuit, we can use the final qubit placements as the initial qubit mapping for our target compilation. This method ensures that qubits needing early interaction in the actual circuit are optimally positioned close together within the QPU’s physical qubit memories (or within a proximity of an EPR pair). This triple compilation strategy, by reusing the same compiler, simplifies the challenging task of the initial qubit mapping.

\textbf{Noise:} We are using model-free policies, and we do not need to know the probability for successful entanglement to optimize the compilation process. With the same argument, we note that although we optimize the time needed for the execution of a quantum circuit, one could change the objective and environment in a way to consider heterogeneous quantum noise in the system. The new model can, for example, maximize the end fidelity of the qubits in the DQC environment given the different CNOT errors (see Figure~\ref{fig:qpu_architecture_guadalupe_16_falcon}). Nevertheless, to be able to do that, the RL environment should be able to simulate such a noise.

\textbf{Quantum switch \& repeater scheduling:} The developed compiler model considers QPUs interconnected via quantum links. However, depending on the DQC architecture and the distance between the QPUs, there might be configurations where quantum switches or repeaters are employed instead \cite{vasantam2022throughput, vardoyan2019stochastic, dai2021entanglement, promponas2024maximizing}. These devices connect with QPUs to facilitate end-to-end entanglement, which is then used for quantum teleportation as usual. To establish this entanglement, the repeaters or switches perform a Bell State Measurement (BSM) on two EPR pairs, each linked to one of the target QPUs. This BSM can be effectively seen as a qubit teleportation of an EPR half to one of the QPUs. Thus, within our model, repeaters or switches can be viewed as specialized QPUs that the initial qubit mapping does not assign logical qubits from the circuit. While existing literature often models arrivals through a stochastic process with a well-defined rate \cite{vasantam2022throughput, vardoyan2019stochastic, dai2021entanglement, promponas2024maximizing, fittipaldi2023linear}, our model adapts to actual requests from the compiler for implementing remote gates from actual quantum circuits. This adjustment is well-suited to the demands of near-term NISQ devices, where the specifics of the circuit and the scheduler’s decisions should not be based on average behavior but on specific case requirements.

\textbf{Tokenize a circuit:} In the RL model developed in Section~\ref{sec:reinforcement_learning_implementation}, our quantum compiler is trained to efficiently handle circuits up to $G_{max}$ gates. To extend this capability to larger circuits, we propose a method akin to the "lookahead" technique used in classical compilers. This method involves tokenizing the circuit into manageable blocks. Each block is compiled independently, where the final qubit mapping of one block serves as the initial qubit mapping for the subsequent block. While this approach may compromise some degree of optimality, it enables the compilation of more complex circuits beyond the original capacity of our compiler, thereby enhancing its practical utility and flexibility.

\textbf{Quantum teleportation inside a single QPU:} In our model of the DQC environment, we utilize the generation of EPR pairs to facilitate gate and qubit teleportation between distant QPUs. However, teleportation may also prove beneficial within a single, sufficiently large QPU for executing infrequent gates between spatially distant qubits. This model allows us to explore the advantages of implementing teleportation within a single QPU. Notably, generating an EPR pair within a single QPU is simpler than between distant QPUs, as it only involves a Hadamard gate followed by a CNOT gate, without the need for fiber optics or flying qubits.

\textbf{Add remote operations in action space:} Note that although the action space developed in Section~\ref{ssec:DQC_compiler} was based on qubit and gate teleportation for remote operations, one could extend the model to use cat-entanglements as a different action by appropriately extending the analysis \cite{andres2019automated, g2021efficient, sundaram2022distribution}.

\textbf{Unary and three-qubit gates:} Our model, which currently relies on quantum circuits with only CNOT gates, can be straightforwardly extended to incorporate unary gates and three-qubit gates such as Toffoli gates by simply modifying the DAG.

\section{Numerical Results}
\label{sec:numerical_results}

In this section, we demonstrate the effectiveness of the RL-based compiler model introduced in the paper, specifically its capability to reduce the expected completion time. The code developed for these simulations is open-sourced.\footnote{https://github.com/ppromponas/CompilerDQC.git} We conducted experiments in a DQC environment with two QPUs as shown in Figure~\ref{fig:qpu_architecture_guadalupe_16_falcon} connected with a quantum link. The training involved random circuits with CNOT gates randomly generated between logical qubits. For every episode, we generate a new random circuit to be compiled and the qubits are initially placed randomly in the QPUs. We used the following reward shaping parameters: $R_{score} = 500$, $R_{success} = -3000$, $R_{fail} = 3000$, and $R_{stop} = -20$. The weight for the quantum link was set to $w_{qlink} = 30$, with $\xi = 18$. The cooldown parameters for "tele-qubit" and "tele-gate" were set at $\kappa = \lambda = 5$. We set the deadline $N = 1500$. Hence the compiler has 1500 time slots before qubit decoherence. This signifies that the compiler can select  "stop" up to 1500 times before the qubits fully decohere and it would correspond to $1500$ matchings/actions in the MDP formulation (see Section~\ref{ssec:DQC_compiler}). 
Finally, we set $|Q| = 18$ for the random circuits, emphasizing that a single QPU considered is unable to execute such circuits.


To identify the optimal learning agent for our MDP, we evaluated various RL methods and fine-tuned the hyperparameters through extensive experimentation. For our best performing method, DDQN, the following set of hyperparameters were found to be efficient for most of the compiler configurations: learning rate $lr=0.00001$, batch size for training the learning agent $\beta=2560$, memory buffer for learning $BS=100000$, epsilon decay denominator $\epsilon_d=50$, discount rate for RL $\lambda=0.99$ and the target network update parameter $\tau=0.001$ (for DDQN). The learning agent was scheduled for training after every $5$ actions, with each training session comprising $10$ iterations of training steps. The neural network architecture that was used (unless otherwise mentioned) consisted of a multi-layer perceptron with two hidden layers (Hadamard product with $S_{msk}$ is not counted as hidden layer), the first having $140$ neurons and the second having $150$ neurons. A ReLU activation layer followed the second hidden layer to ensure all outputs were non-negative to ensure that masks correctly adjust the Q-value for infeasible actions to zero and maintain non-negative real numbers for others. 

\subsection{Reinforcement Learning Algorithms}
\label{ssec:num_learning_methods}

\begin{figure}[!t]
\centering
\hspace{-0.5cm}
\subfloat[\quad Cumulative Reward]{ 
\vspace{-0.3cm}
\includegraphics[height=1.7in]{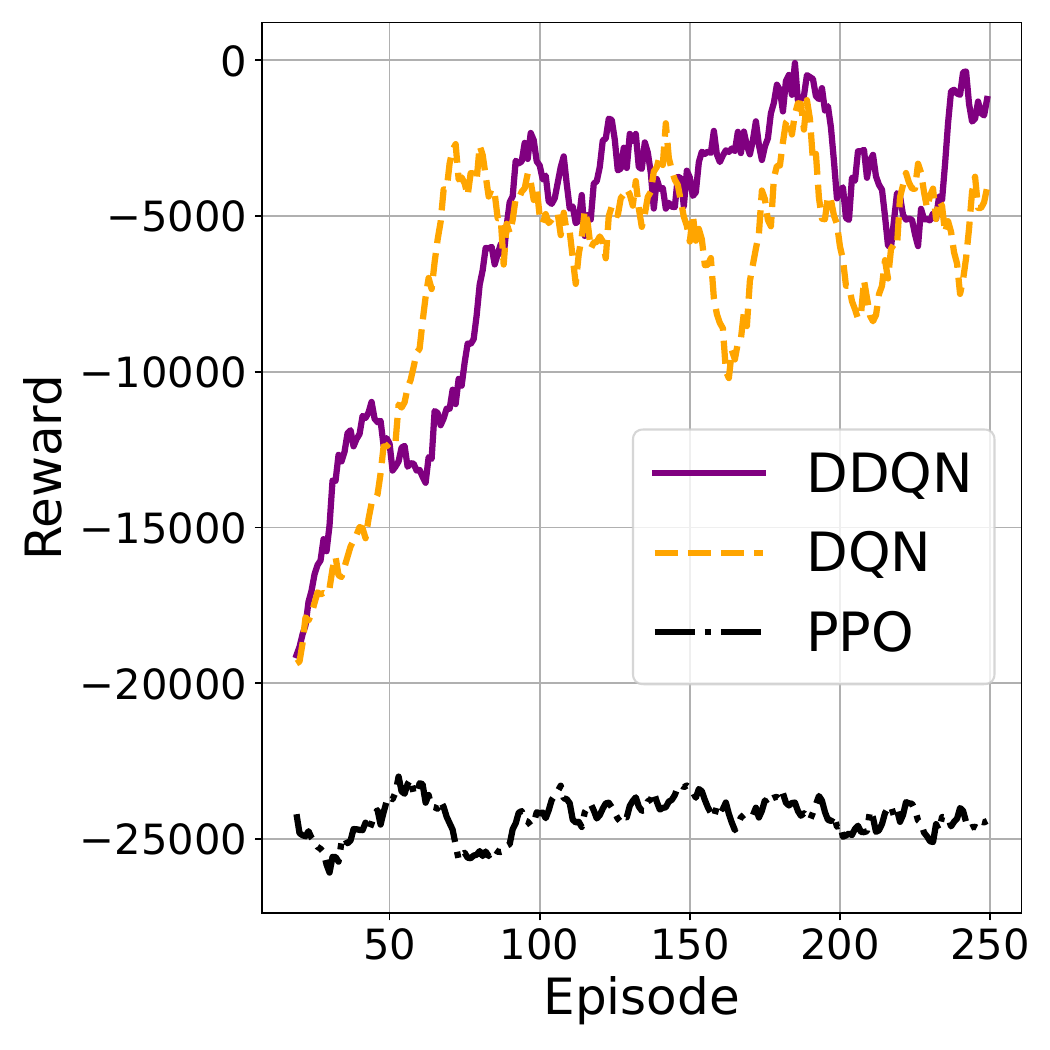}}
\subfloat[\quad Time Elapsed]{
\vspace{-0.3cm}
\includegraphics[height=1.7in]{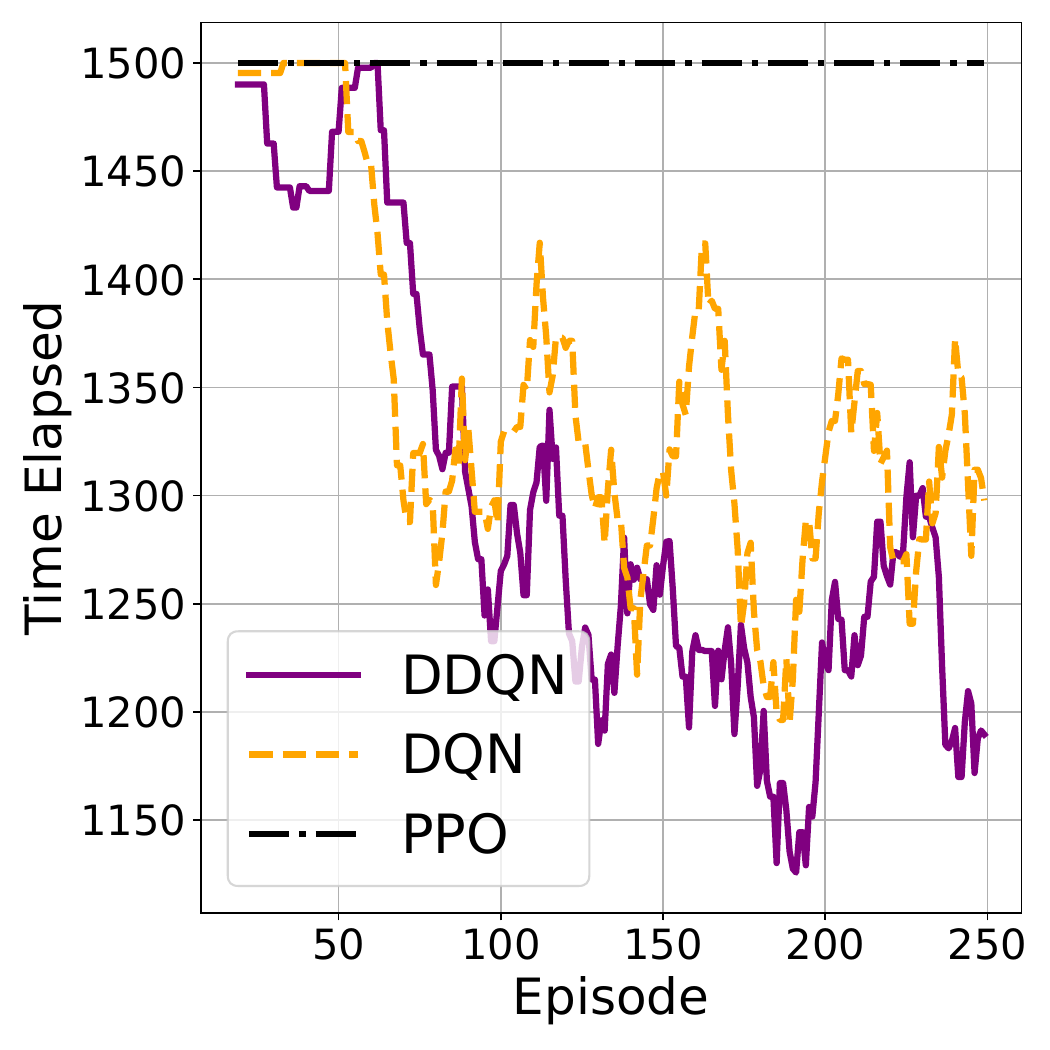}}
\vspace{-2mm}
\caption{Time evolution of (a) cumulative reward and (b) time elapsed until successful quantum circuit compilation or deadline expiry, for DDQN, DQN, and PPO.}
\label{fig:rl_method}
\vspace{-2mm}
\end{figure}

In this experiment, we evaluate the performance of three RL architectures: DQN, DDQN, and PPO, using random circuits with 30 gates and probability for successful entanglement $p_{gen} = 0.95$. Figure~\ref{fig:rl_method}(a) displays the time evolution of the reward across episodes. Notably, PPO fails to increase the reward, whereas DQN and DDQN progressively learn to optimize it. To illustrate the correlation between reward and the compiler's efficiency, Figure~\ref{fig:rl_method}(b) tracks the actual time elapsed (counted by the number of "stops") until the compiler either successfully compiles the circuit or reaches the deadline. PPO never successfully compiled the circuit, while DQN and DDQN initially struggled but eventually learned to compile the circuits more efficiently, thereby reducing the compilation time. This demonstrates that the reward shaping, detailed in Section~\ref{sec:reinforcement_learning_implementation}, effectively links higher rewards to reduced compilation time. Furthermore, the time elapsed corresponds to the depth of the compiled circuit, based on the DQC environment’s time slot definition. In case of $p_{gen} = 1$ the depth of the compiled circuit coincides with the time elapsed until the compilation of the circuit. We note that DDQN performed marginally better than the other value-based approach, DQN, so we only consider DDQN for the rest of the section.

\subsection{Varying Probabilities of Successful Entanglement}
\label{ssec:num_probability_entanglement}

\begin{figure}[!t]
\centering
\hspace{-0.5cm}
\subfloat[\quad Cumulative Reward]{ 
\vspace{-0.3cm}
\includegraphics[height=1.7in]{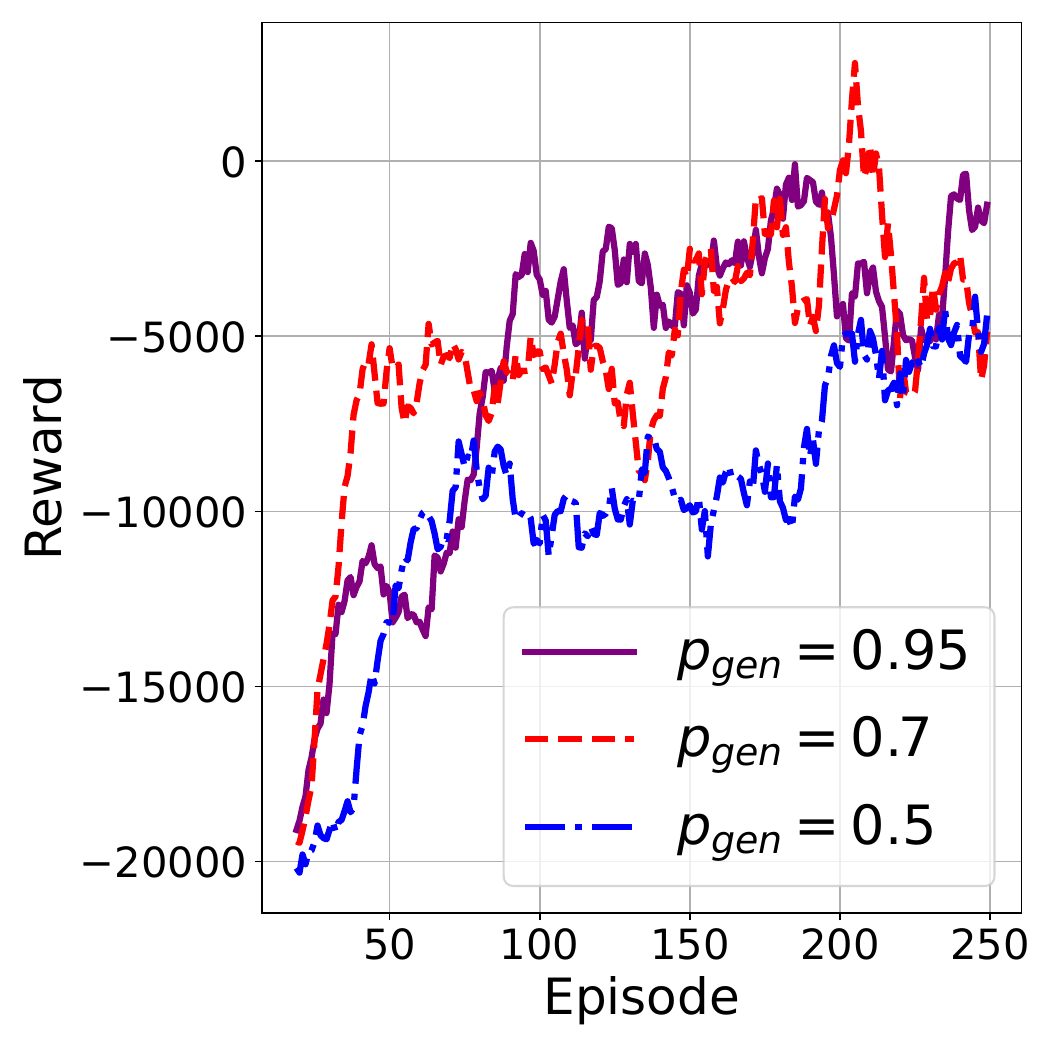}}
\subfloat[\quad Time Elapsed]{
\vspace{-0.3cm}
\includegraphics[height=1.7in]{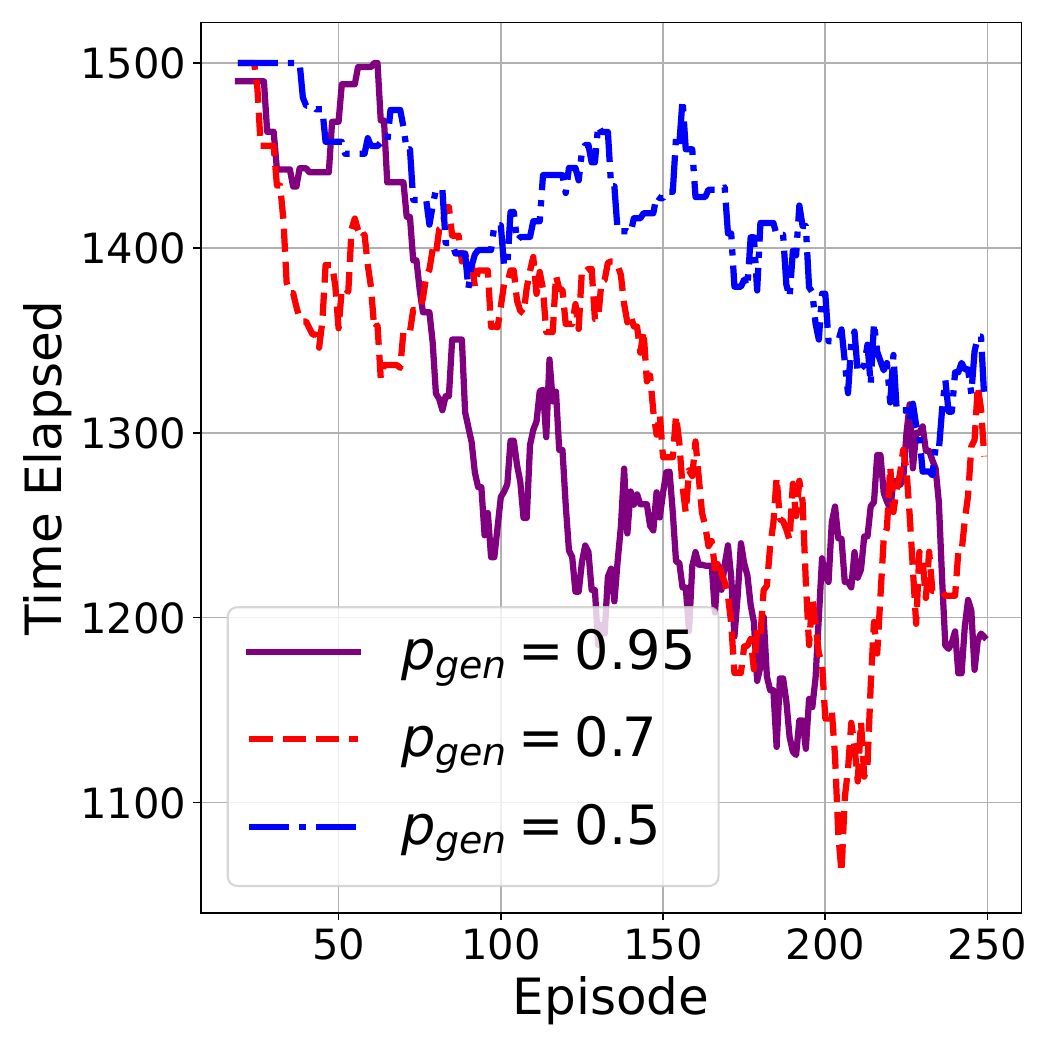}}
\vspace{-2mm}
\caption{Time evolution of (a) cumulative reward and (b) time elapsed until successful quantum circuit compilation or deadline expiry, for various probabilities for successful EPR generation $p_{gen} \in \{0.95, 0.7, 0.5\}$.}
\label{fig:prob_entanglement}
\vspace{-2mm}
\end{figure}

In this experiment we vary the probability for successfully generating an EPR pair, $p_{gen}$. We study the cases of $p_{gen} \in \{0.95, 0.7, 0.5\}$ and we plot in Figure~\ref{fig:prob_entanglement} the time evolution of (a) the cumulative reward and (b) time elapsed until successful quantum circuit compilation or deadline expiry (depending on what happened first). In this experiment, to boost the performance of the compiler, we had to increase the neural network to 240 and 200 neurons for the hidden layers for the cases of $p_{gen} = 0.7$ and $p_{gen} = 0.5$. Observe that DDQN was able to learn how to optimize the reward (Figure~\ref{fig:prob_entanglement}(a)) and compile the circuits successfully (Figure~\ref{fig:prob_entanglement}(b)) even when we increase the uncertainty in the model.

\subsection{Varying Number of Gates}
\label{ssec:num_gates}

Recall that we arbitrarily set the deadline to $N=1500$. This experiment studies what number of gates we can compile in this deadline. For that reason, we test random circuits with  $30, 40$ and $50$ gates for the training of the RL agent. However, a trained compiler could compile circuits with more gates by partitioning the circuit into blocks, where the final configuration of each block serves as the initial qubit mapping for compiling the subsequent block.

\begin{figure}[!t]
\centering
\hspace{-0.5cm}
\subfloat[\quad Cumulative Reward]{ 
\vspace{-0.3cm}
\includegraphics[height=1.7in]{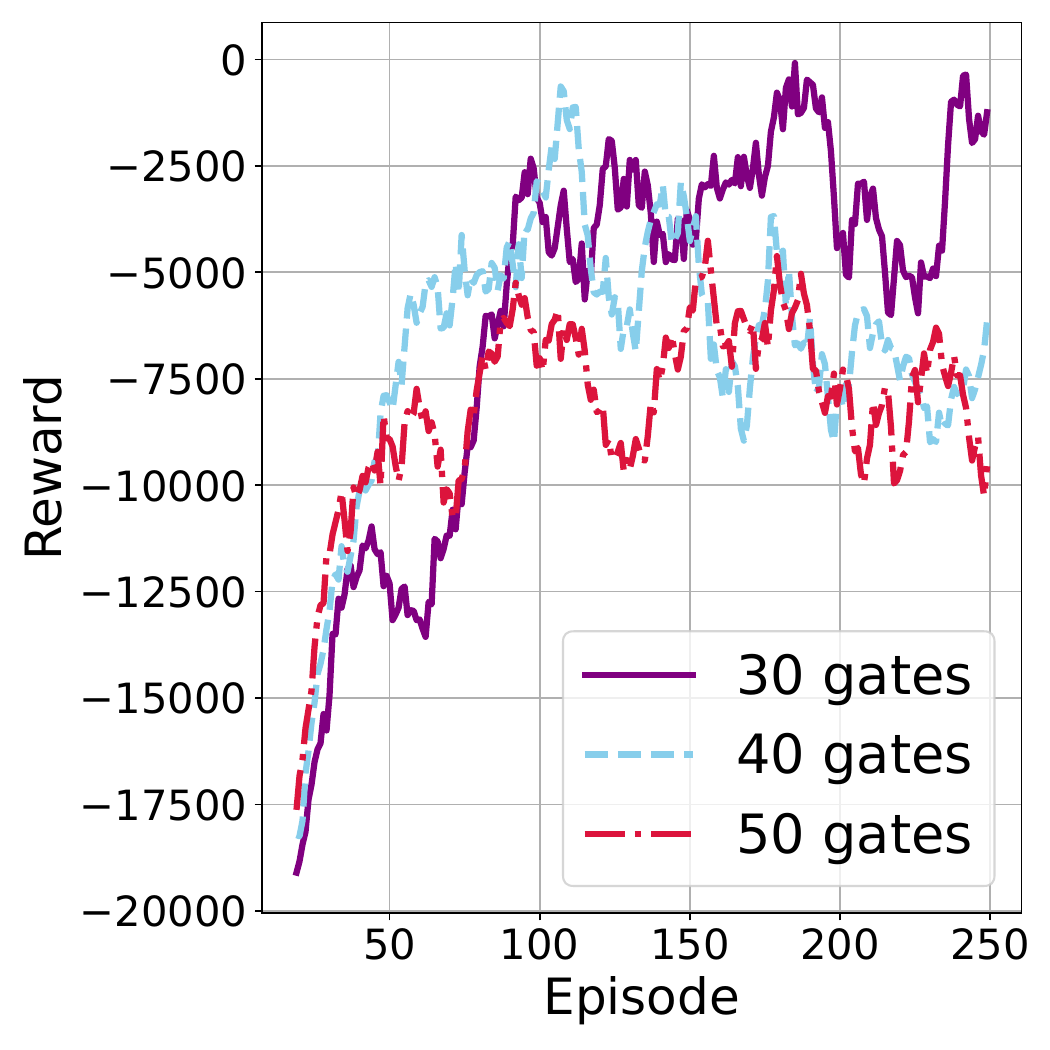}}
\subfloat[\quad Time Elapsed]{
\vspace{-0.3cm}
\includegraphics[height=1.7in]{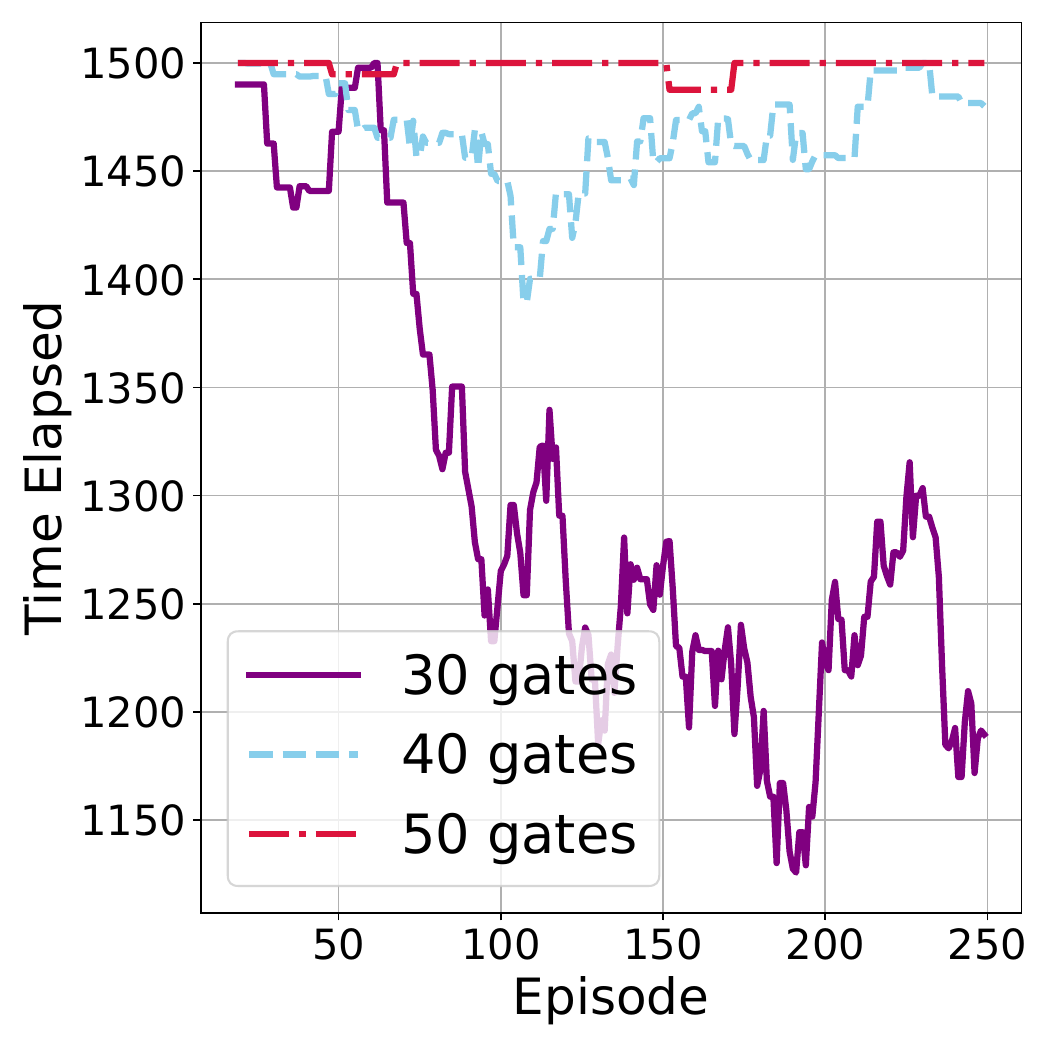}}
\vspace{-2mm}
\caption{Time evolution of (a) cumulative reward and (b) time elapsed until successful quantum circuit compilation or deadline expiry, for random circuits comprising of $30, 40$ and $50$ CNOT gates.}
\label{fig:gates}
\vspace{-2mm}
\end{figure}

Observe from Figure~\ref{fig:gates}(a) that in all of the cases the compiler was able to learn how to increase the reward - which corresponds to learning how to complete more and more gates in the circuit. However, as indicated in Figure~\ref{fig:gates}(b), the deadline proved insufficient for the successful compilation of circuits with $50$ gates. For circuits with $40$ gates, the deadline barely allowed the compiler to sometimes complete the compilation successfully. Nevertheless, the inconsistency in achieving this led to a failure in maintaining increased rewards when compilations were unsuccessful.

\section{Conclusion}

We introduce a novel compiler for DQC that, unlike existing approaches,  prioritizes reducing the expected execution time by jointly managing the generation and routing of EPR pairs, scheduling remote operations, and injecting SWAP gates to facilitate the execution of local gates. This compiler can be employed to jointly optimize the entanglement distribution network and the qubit routing of the logical qubits of the quantum circuit to successfully compile and execute the latter. It aims to facilitate successful compilation and execution, particularly in the near term when resources are limited, by providing personalized execution and resource allocation tailored to each DQC environment and quantum circuit. In the future, we plan to broaden the scope of our simulation studies by testing the compiler under various scenarios.

\section{Acknowledgements}
The research work was supported by the Army Research Office MURI under the project number W911NF2110325 and by the National Science Foundation under project numbers EEC-1941583 CQN ERC and CNS 1955744. The authors would also like to thank Richard Chen, Oliver Crampton, and Dionysis Kalogerias for their feedback and recommendations.

    
    

\bibliographystyle{IEEEtran}
\bibliography{references}

\end{document}